\documentclass[showpacs,amsmath,superscriptaddress,aps,reprint]{revtex4-2} 

\usepackage{graphicx}
\usepackage{hyperref}
\usepackage{amsmath} 
\usepackage{dcolumn}
\usepackage{bm}

\usepackage[utf8]{inputenc}
\usepackage[T1]{fontenc}
\usepackage{mathptmx}

\begin{document}

\title
{Self-aligned hybrid nanocavities using atomically thin materials}

\author{C.~F.~Fong}
\email[Corresponding author: ]{cheefai.fong@riken.jp}
\affiliation{Nanoscale Quantum Photonics Laboratory, RIKEN Cluster for Pioneering Research, Saitama 351-0198, Japan}

\author{D.~Yamashita}
\affiliation{Quantum Optoelectronics Research Team, RIKEN Center for Advanced Photonics, Saitama 351-0198, Japan}
\affiliation{Platform Photonics Research Center, National Institute of Advanced Industrial Science and Technology (AIST), Ibaraki, 305-8568, Japan}

\author{N.~Fang}
\affiliation{Nanoscale Quantum Photonics Laboratory, RIKEN Cluster for Pioneering Research, Saitama 351-0198, Japan}

\author{S.~Fujii}
\affiliation{Quantum Optoelectronics Research Team, RIKEN Center for Advanced Photonics, Saitama 351-0198, Japan}
\affiliation{Department of Physics, Faculty of Science and Technology, Keio University, Yokohama, 223-8522, Japan}

\author{Y.-R.~Chang}
\affiliation{Nanoscale Quantum Photonics Laboratory, RIKEN Cluster for Pioneering Research, Saitama 351-0198, Japan}

\author{T.~Taniguchi}
\affiliation{Research Center for Materials Nanoarchitectonics, National Institute for Materials Science, Tsukuba 305-0044, Japan}

\author{K. Watanabe}
\affiliation{Research Center for Electronic and Optical Materials, National Institute for Materials Science, Tsukuba 305-0044, Japan}

\author{Y.~K.~Kato}
\email[Corresponding author: ]{yuichiro.kato@riken.jp}
\affiliation{Nanoscale Quantum Photonics Laboratory, RIKEN Cluster for Pioneering Research, Saitama 351-0198, Japan}
\affiliation{Quantum Optoelectronics Research Team, RIKEN Center for Advanced Photonics, Saitama 351-0198, Japan}

\begin{abstract}
Two-dimensional (2D) van der Waals layered materials with intriguing properties are increasingly being adopted in hybrid photonics. The 2D materials are often integrated with photonic structures including cavities to enhance light-matter coupling, providing additional control and functionality. The 2D materials, however, needs to be precisely placed on the photonic cavities. Furthermore, the transfer of 2D materials onto the cavities could degrade the cavity quality $(Q)$ factor. Instead of using prefabricated PhC nanocavities, we demonstrate a novel approach to form a hybrid nanocavity by partially covering a PhC waveguide post-fabrication with a suitably-sized 2D material flake. We successfully fabricated such hybrid nanocavity devices with hBN, WSe$_2$ and MoTe$_2$ flakes on silicon PhC waveguides, obtaining $Q$ factors as high as $4.0\times10^5$. Remarkably, even mono- and few-layer flakes can provide sufficient local refractive index modulation to induce nanocavity formation. Since the 2D material is spatially self-aligned to the nanocavity, we have also managed to observe cavity PL enhancement in a MoTe$_2$ hybrid cavity device, with a cavity Purcell enhancement factor of about 15. Our results highlights the prospect of using such 2D materials-induced PhC nanocavity to realize a wide range of photonic components for hybrid devices and integrated photonic circuits. 
\end{abstract}

\maketitle

\section{Main}\label{sec:MAIN}

Two-dimensional (2D) van der Waals layered materials such as graphene, hexagonal boron nitride (hBN) and transition metal dichalcogenides (TMDC) are garnering significant attention for both fundamental science and device applications~\cite{Xu2014,krasnok2018,liu2019a,he2021}. In particular, semiconducting TMDCs have a direct bandgap at monolayer thickness~\cite{mak2010,ruppert2014}. They exhibit bright optical emission which is governed by their exciton (coulomb-bound electron-hole pair) responses even up to room temperature due to the large exciton binding energies of hundreds of meV~\cite{zhu2015a}. Many 2D materials exhibit large optical nonlinearities~\cite{autere2018}, which could also be intricately linked to the valley polarization~\cite{seyler2015,zhang2020b}. The amenability to strain and defect engineering also makes 2D materials promising for the creation of emitters for quantum light sources~\cite{liu2019a,turunen2022,hotger2023}. 2D materials can also be stacked to achieve new functionalities or physical phenomena. For example, mono- and few-layer TMDCs are often used together with hBN as the insulating layer and with graphene for gate tunable functionalities~\cite{cadiz2017,martin2020}. In addition, stacks of 2D material layers could give rise to proximity effects~\cite{Avsar2017b,seyler2018a,zhong2020a}, as well as moiré related behaviour~\cite{seyler2019,baek2020,zhang2020c,shinokita2022}. 

The integration of two-dimensional (2D) materials with nanophotonic architectures~\cite{gu2012,ono2020,maiti2020,fang2022,maggiolini2023}, offers a promising avenue for tailoring the dielectric environment and local optical density of states that govern the interactions of 2D excitons with light. This strategy enables the manipulation of light-matter coupling, thereby facilitating the realization of practical hybrid devices such as lasers, nonlinear sources, quantum emitters and photodetectors. To achieve optimal device performance, it is imperative to couple 2D materials with cavities exhibiting a high quality factor over mode volume ratio ($Q/V_\text{mode}$). The planar air hole photonic crystal (PhC) nanocavities~\cite{song2005,nomura2010,kuruma2018}, promise strong light confining in an ultrasmall mode volume, rendering them highly attractive for the aforementioned hybrid device applications. 

In previous reports, 2D materials such as TMDC mono- and few-layer flakes are usually transferred onto prefabricated nanocavities. The flake transfer requires precise alignment and placement to the cavity. Aside from absorption by the flake, the transfer of a TMDC flake changes the dielectric environment of the cavity and could induces other optical losses which results in significant degradation in the cavity quality $(Q)$ factor, limiting device performance~\cite{majumdar2013,wu2014,fryett2016,ota2017,li2021b}. This effect becomes even more severe and unpredictable with stacks of 2D materials with irregular shapes and thicknesses. Instead of being a detriment, the change in the dielectric environment caused by the 2D material can be utilized to enable alternative methods for cavity light-matter coupling. 

\begin{figure*}
\includegraphics[width=1.8 \columnwidth]{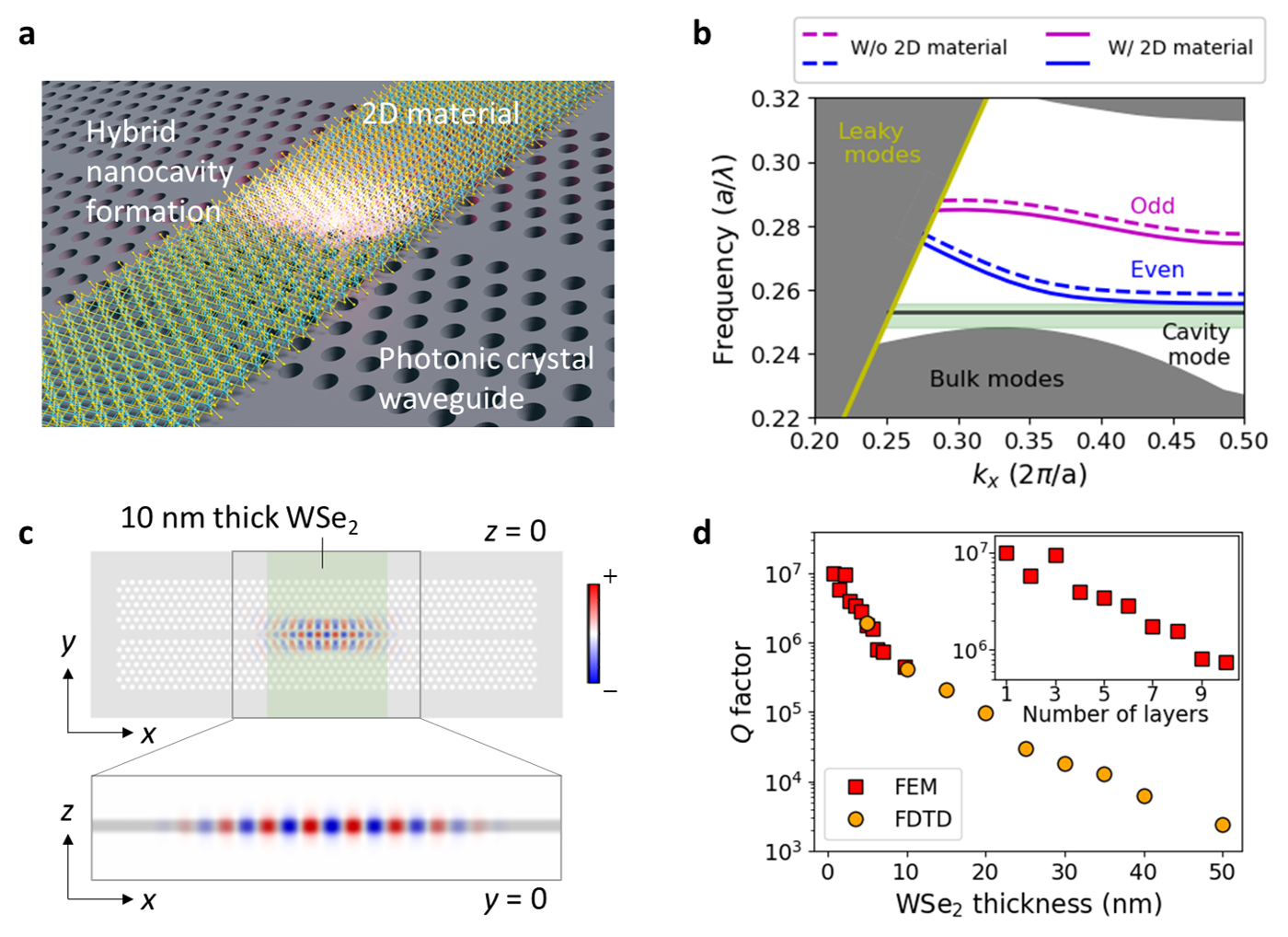}
\caption{
\label{Fig1}\textbf{2D material hybrid PhC mode gap nanocavity.} \textbf{a} Schematic of a device structure showing cavity formation at the location of the 2D material.  \textbf{b} Photonic bands of the PhC waveguide showing the odd and even guided modes as labelled, as well as the mode gap in green and the cavity mode within. The solid (dashed) lines represent the guided modes with (without) the 2D material. \textbf{c} Electric field E$_y$ profile of the cavity mode in the $xy$- and $xz$-planes. The green shaded region represents the 2D material flake. \textbf{d} The dependence of the cavity $Q$ factor against thickness of WSe$_2$. (Inset) Plot of the $Q$ factor against number of layers. 
}
\end{figure*}

In this work, we demonstrate the formation of hybrid PhC mode gap nanocavities by partially covering PhC waveguides post-fabrication with suitably-sized 2D material flakes. The presence of the flake modulates the local refractive index, causing a mode frequency mismatch, leading to optical confinement and thus cavity formation. The flake is spatially self-aligned to the cavity, facilitating optical coupling. We have successfully fabricated such nanocavities with various 2D materials including hexagonal boron nitride (hBN), tungsten diselenide (WSe$_2$) and molybdenum ditelluride (MoTe$_2$), achieving high $Q$ factors of $10^4$--$10^5$. Contrary to our initial expectations, we have discovered that even a monolayer flake could give rise to sufficient local refractive index modulation to form a hybrid nanocavity. In fact, according to simulation results, the thinner the flake, the more moderate the refractive index modulation, the higher the $Q$ factor. In such hybrid systems, the monolayer represents the extreme limit of index modulation, promising high $Q$ factors. We have further observed the self-aligned coupling of MoTe$_2$ photoluminescence (PL) in a hybrid nanocavity device, giving rise to Purcell enhanced emission and lifetime reduction corresponding to a Purcell factor of about 15.

\section{Design of hybrid nanocavity}\label{sec:DESIGN}

For simulating our device structure, we employ the air-suspended W1 line defect PhC waveguide~\cite{notomi2008a} made of silicon (refractive index, $n_{Si}$ = 3.48), consisting of a triangular array of air holes with lattice period $a$, with a 2D material flake covering a section of the waveguide (Fig.~\ref{Fig1}a). We typically consider a PhC waveguide with 48 and 14 air holes along the $\Gamma$--$K$ and $\Gamma$--$M$ directions, respectively, with the air hole radius, $r$ = 0.28$a$ and $a$ = 340 nm. The PhC slab thickness is 200 nm. The photonic band structure of the transverse-electric-like (TE-like) modes of the PhC waveguide is shown in Fig.~\ref{Fig1}b. There are two guided modes, referred to as odd and even in accordance with the symmetry of the $E_y$ field distribution about the $x$-axis. 

When a 2D material flake is present on the PhC waveguide, the local effective refractive index increases, red shifting the frequencies of the guided modes. The frequency mismatch between the regions with and without the 2D material flake gives rise to field confinement and thus the formation of a cavity. There are corresponding cavity modes to each of the guided modes with the mode resonances in the near infrared (NIR) regime.  The frequencies of the hybrid nanocavity modes are usually lower than the band edge (frequency at $k_x = 0.5$) of the corresponding guided modes. The even cavity modes exhibit much higher $Q$ factors as the modes are within the mode gap --- the frequency range between the even guided mode edge and the lower edge of the photonic band gap (green region in Fig.~\ref{Fig1}b). In this work, we will mainly focus on the even cavity modes. 

\begin{figure*}[t]
\includegraphics[width=2.0 \columnwidth]{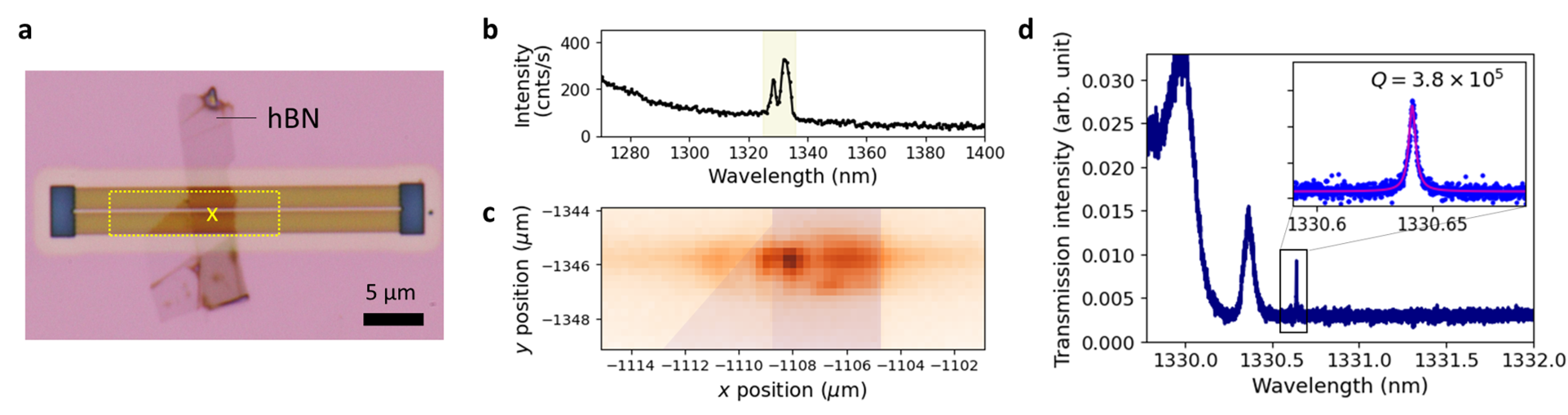}
\caption{
\label{Fig2}\textbf{hBN-induced hybrid nanocavity.} \textbf{a} Optical micrograph of an hBN hybrid nanocavity device. The scale bar represents 5 µm. The “x” symbol marks the location of the PL measurement in (b). The dotted box outlines the area of the 2D imaging shown in (c). \textbf{b} PL spectrum showing signatures of cavity emission. \textbf{c} 2D image of the integrated PL intensity over the wavelength range of 1325--1336 nm showing localized cavity emission at the hBN flake (the flake is indicated by the darkened region). \textbf{d} Transmission spectra showing the cavity peak. (Inset) A closer look at the narrow peak fitted to a Lorentzian function (magenta line).
}
\end{figure*}

To investigate the cavity mode properties, we consider a rectangular WSe$_2$ flake (refractive index, $n_{WSe_2}$ = 3.95~\cite{laturia2018}) of 10 nm thickness with a lateral width of 14$a$, partially covering the surface of the PhC waveguide. The flake is assumed to cover the PhC structure completely along the $y$-direction. The simulated $E_y$ field amplitude profile of the fundamental even cavity mode is shown in Fig.~\ref{Fig1}c. Due to the minimal thickness of the 2D material flake, most of the field is concentrated within the slab in the region covered by the WSe$_2$ flake. 

The simulated fundamental cavity mode $Q$ factor dependence on the thickness of the WSe$_2$ flake is summarized in Fig.~\ref{Fig1}d. At a thickness of 30 nm, the cavity $Q$ is about 10$^4$. As the thickness of the WSe$_2$ flake decreases, the $Q$ factor increases, reaching a theoretical value as high as 10$^6$ for few-layer flakes. Simulation results indicate that not only can a monolayer form a cavity, but it also promises an ultrahigh $Q$ factor of the order of 10$^7$. The atomically thin nature of the flake causes a less abrupt change of the refractive index in space, leading to a “gentle” perturbation to the fields. This in turn corresponds to having less field components within the leaky region in the momentum space (see Supplementary Fig. S1), indicating less loss via coupling to free space, and thus results in ultrahigh $Q$ factors. 

Other properties such as the mode profile remains similar for the range of thickness considered in our simulations. The cavity mode wavelength increases with flake thickness while the cavity mode volume remains relatively constant before increasing when the flake thickness decreases below 10 nm (Supplementary Fig. S1). These trends also apply to other 2D materials. Further details about other factors that affect the cavity $Q$ factor such as the flake lateral width and refractive index are provided in Supplementary Fig.~S2.

\section{\texorpdfstring{\MakeLowercase{h}BN}{hBN}-induced hybrid \texorpdfstring{P\MakeLowercase{h}C}{PhC} nanocavity}\label{sec:hBN}

Dielectric hBN --- which is often used to encapsulate TMDC to prevent exposure to air --- is also a suitable 2D material to form hybrid nanocavities. In particular, high crystal quality hBN flakes with low defect densities promise high $Q$ factors. Figure~\ref{Fig2}a shows an optical micrograph of a fabricated hBN-on-PhC waveguide device. The PhC slab thickness is 200 nm, and the waveguide consists of 48 and 14 air holes along the $\Gamma$--$K$ and $\Gamma$--$M$ directions, respectively. The lattice and the air hole radius are $a$ = 360 nm and $r$ = 0.27$a$, respectively. For this PhC waveguide, semi-circular output couplers are included at each end of the waveguide to facilitate laser transmission measurements. The hBN flake is transferred onto the PhC waveguide post-fabrication (see Methods section for further details). The thickness of the hBN flake is estimated to be about 20 nm based on the optical contrast, and its lateral width measured along the waveguide is about 12$a$.

By optically exciting the silicon PhC device above the bandgap, the weak emission from silicon substrate can couple to the guided and/or cavity modes, manifesting as peaks in the PL spectra. For this particular sample, emission peaks appear near the frequency of the even guided mode edge (Fig.~\ref{Fig2}b). By performing 2D PL imaging, we confirm that the emission peaks only appear when exciting at the hBN flake along the waveguide, further indicating the formation of a cavity (Fig.~\ref{Fig2}c). The polarization properties of the peak are broadly consistent with expectations in accordance with the mode profile (see Supplementary Fig. S3). However, the actual linewidth of the peaks, and thus the $Q$ factors, cannot be determined using the spectrometer due to insufficient resolution. 

We then carried out transmission measurements by exciting at the hBN flake and detecting the scattered light from the outcoupler to right of the waveguide. A few features can be seen in the transmission spectrum (Fig.~\ref{Fig2}d): the guided mode edge can be seen at about 1330.00~nm, as well as a relatively broad peak centered at 1330.34 nm and a narrow peak at 1330.61~nm. The broad peak has a linewidth of 60.2 pm, corresponding to $Q = 2.2\times10^4$ while the narrow peak has a linewidth of 3.54 pm, giving $Q = 3.8\times10^5$. Given the relative position and the $Q$ factors of the peaks, the narrow and broad peaks should correspond to the fundamental and the first higher order cavity modes.

\begin{figure*}[t]
\includegraphics[width=2.0 \columnwidth]{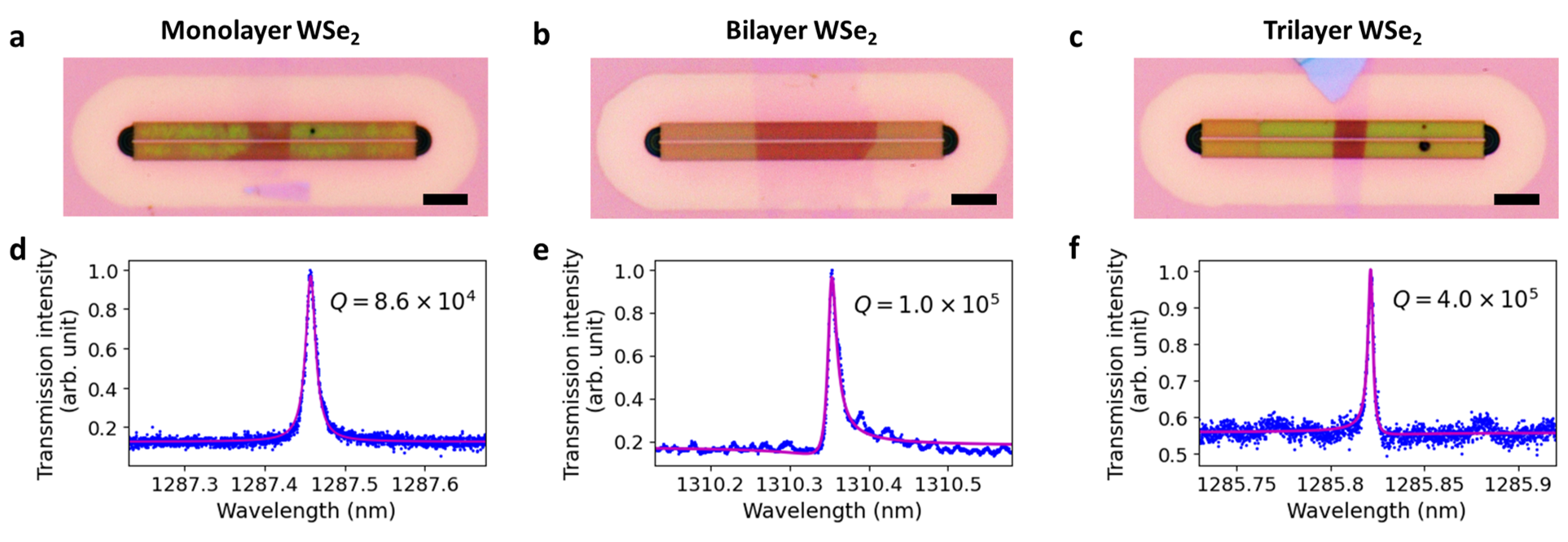}
\caption{
\label{Fig3}\textbf{Mono- \& few-layer WSe$_2$ hybrid nanocavity.} \textbf{a-c}  Optical micrograph of mono-, bi- and trilayer WSe$_2$-induced nanocavity device. The waveguide consists of 96 and 14 air holes along the $\Gamma$ -- $K$ and $\Gamma$ -- $M$ directions, respectively, to facilitate ease of 2D material flake transfer. The brightness and contrast of the images have been adjusted to improve the visibility of the WSe$_2$ flake. For the mono- and trilayer device, the PhC waveguide has $a$ = 348 nm while the bilayer device is of $a$ = 356 nm. The lateral widths along the waveguide of the monolayer, bilayer and trilayer flakes of the devices are 13$a$, 36.5$a$ and 10$a$, respectively. Some residues could be left on the sample during the transfer process in the monolayer device. However, there is no significant effect on the transmission spectra. \textbf{d-f} the corresponding cavity peak in the transmission spectra for each of the devices. The transmission measurements are performed by exciting the left output coupler and detecting at the right output coupler for the mono- and trilayer devices. As for the bilayer device, the laser excitation is focused on the sample and the transmitted signal is detected at the right output coupler. The peak in (d) is fitted with the Lorentzian function, while the peaks in (e) and (f) are fitted with the Fano resonance. The scale bars in (a-c) represents 5 $\mu$m.
}
\end{figure*}

The extracted $Q$ factors from the resonant peaks of the transmission spectra are the so-called loaded $Q$ factors which consist of the intrinsic cavity $Q$ ($Q_\text{i}$) and cavity-waveguide coupling $Q$ ($Q_\text{c}$)~\cite{akahane2005}: $\frac{1}{Q_{\text{l}}} = \frac{1}{Q_{\text{i}}} + \frac{1}{Q_{\text{c}}}$. For this hBN device, by changing the transmission measurement configuration to excite at the left output coupler and detecting light from the right output coupler, we are able to observe the resonant dip that corresponds to the absorption of light by the cavity, allowing us to calculate $Q_\text{i}$ of this device to be about 4.5$\times10^5$ (Supplementary Fig. S4). 

Overall, our results show that the flake and the nanocavity are colocalized, essentially forming a self-aligned device. The enhanced Si PL emission also indicates cavity light-matter coupling.

\section{Hybrid \texorpdfstring{P\MakeLowercase{h}C}{PhC} nanocavities using atomically thin \texorpdfstring{WS\MakeLowercase{e}}{WSe}$_2$}\label{sec:WSe2}

Compared to hBN, it is easier to obtain and identify mono- and few-layer flakes with TMDCs. Motivated by our simulation results, we conduct experiments using WSe$_2$ as it is relatively stable in air. We successfully fabricate seven hybrid PhC nanocavity devices using monolayer to trilayer flakes. The discussion here will focus on three exemplary devices with monolayer, bilayer and trilayer flakes  (Fig.~\ref{Fig3}). Summaries of all device configurations and $Q$ factors are presented in Fig. S10 and Supplementary Table 1. 

The transmission spectrum for each device is shown in Fig.~\ref{Fig3}e-f. The cavity peaks appear at the expected wavelengths in accordance with the PhC waveguide lattice period. Due to the presence of background transmission signal, its interference with the cavity peak could results in an asymmetric line shape. By fitting the peaks to either the Lorentzian (symmetric line shape) or Fano resonance (asymmetric line shape), $Q_\text{l}$ are determined to be 8.6$\times10^4$, 1.0$\times10^5$ and 4.0$\times10^5$, for the monolayer, bilayer and trilayer devices, respectively. Only for the bilayer sample are we able to obtain the necessary transmission spectra with the cavity resonant dip and peak in the different measurement configurations to extract the $Q_\text{i}$ to be about 3.7$\times10^5$. The cavity $Q$ factors do not show an obvious increase with the decrease flake thickness even after considering the flake lateral widths. Various loss mechanisms could affect the extracted $Q$ factors; further descriptions of the possible mechanisms are provided in the discussion section. Our results here show unambiguously that even a monolayer flake could give rise to sufficient local refractive index modulation to form a high $Q$ hybrid nanocavity despite being only one-atom thick. 

\begin{figure*}[t]
\includegraphics[width=1.8 \columnwidth]{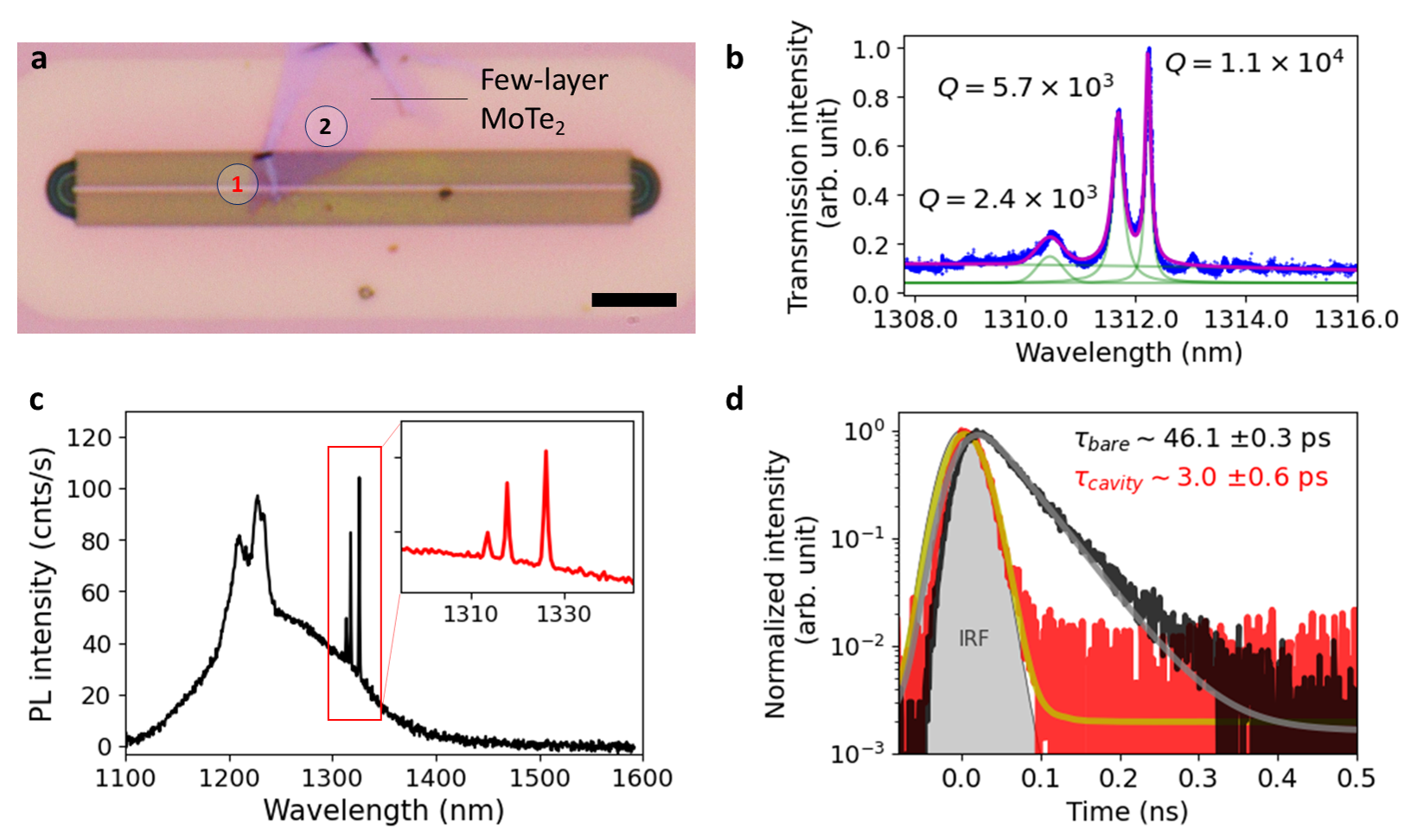}
\caption{
\label{Fig4}\textbf{Purcell enhancement in MoTe$_2$ hybrid nanocavity.} \textbf{a} Optical micrograph of an MoTe$_2$ nanocavity device. The scale bar indicates 5 $\mu$m. \textbf{b} Transmission spectra obtained by exciting at the left and detecting at the right output coupler, showing the cavity peaks. The green curves are the individual Lorentzian fits, and the magenta curve is the cumulative fit. \textbf{c} PL spectrum showing the emission from the MoTe$_2$ and cavity. (Inset) A zoom-in of the region with the cavity emission peak. \textbf{d} Time-resolved PL decay curves of the emission from the cavity and the bare MoTe$_2$ flake (positions 1 \& 2 in (a), respectively) showing the reduced lifetime due to the cavity Purcell effect. The cavity and the bare MoTe$_2$ emission decay curves are fitted using a single and double exponential reconvolution functions, respectively. The grey shaded region shows the IRF.
}
\end{figure*}

\section{Cavity PL enhancement in self-aligned hybrid nanocavity with \texorpdfstring{M\MakeLowercase{o}T\MakeLowercase{e}}{MoTe}$_2$}\label{sec:MoTe2}

MoTe$_2$ is optically active in the NIR, making it an ideal TMDC to induce a nanocavity in the PhC and simultaneously couple its exciton emission to the cavity. Figure~\ref{Fig4}a shows an optical micrograph of a MoTe$_2$ hybrid nanocavity device. The flake is about 6--8 layers thick based on optical contrast and the PL peak position. The lateral width of the flake along the waveguide is about 6$a$. Despite not fully covering the PhC along the vertical ($y$) direction, the flake managed to induce the formation of a cavity. Multiple peaks are visible in the transmission spectrum (Fig.~\ref{Fig4}b) with $Q$ factors of the order of $10^{3}$--$10^{4}$. The lower $Q$ factors are due to a combination of the larger flake refractive index, the small lateral width, as well as the absorption by the flake. Modelling a structure with similar properties as the actual device in FDTD and FEM simulations, we confirmed that the formation of cavity that could support multiple modes is indeed possible (Supplementary Fig. S5).  In fact, a local refractive index modulation of a small area adjacent to the waveguide is sufficient to induce strong optical confinement (Supplementary Fig. S6).  

A broad background from the MoTe$_2$ exciton emission can be seen in the PL spectrum (Fig.~\ref{Fig4}c). Peaks close to 1200~nm correspond to the Fabry-Perot-like emission arising from the odd guided mode, while the narrow peaks at around 1310~nm correspond to the cavity modes. The bright intensity of these peaks indicates that the MoTe$_2$ emission is indeed coupled to and enhanced by the cavity. We further perform time-resolved PL measurements. A comparison of the decay curves of the PL from the cavity and bare MoTe$_2$ (positions 1 \& 2, respectively in Fig.~\ref{Fig4}a) is shown in Fig.~\ref{Fig4}d. From the decay curves, the bare MoTe$_2$ PL lifetime is extracted to be 46.1$\pm$0.3 ps, whereas the cavity emission lifetime is 3.0$\pm$0.6 ps. The cavity decay curve is largely similar to that of the instrument response function (IRF), suggesting that the actual lifetime is shorter than the extracted value. Nonetheless, the PL lifetimes indicate a Purcell enhancement factor,  $F_\text{P}$ of 15$\pm$3. Since the emitter linewidth is broader than the cavity linewidth, we employ the following equation to estimate the theoretical Purcell enhancement factor~\cite{vanexter1996}: $F_\text{P} = 1 + \frac{3\lambda_{\text{cav}}^3}{4\pi^2 n^2} \frac{Q_{\text{emitter}}}{V_{\text{mode}}} \frac{|E(r)|^2}{|E(r)|_{\text{max}}^2}$, where $\lambda_{\text{cav}}$ is the cavity wavelength, $n$ is the refractive index, $Q_\text{emitter}$ is the emitter $Q$ factor, $V_\text{mode}$ is the mode volume and $\frac{|E(r)|^2}{|E(r)|_{\text{max}}^2}$ is the ratio of the field intensity at the location of the emitter and the maximum field intensity of the mode.  Since the MoTe$_2$ emitter is in air, $n$ is taken to be 1. From FEM simulations of a device with a similarly shaped and thick flake, $V_\text{mode} = 0.04 (\lambda_{\text{cav}}/n)^3$ and $\frac{|E(r)|^2}{|E(r)|_{\text{max}}^2} = 0.49$. Based on the linewidth of the MoTe$_2$ PL emission, $Q_\text{emitter} = 9$, resulting in a theoretical $F_\text{P}$ of 9, close to the experimental $F_\text{P}$. Our results unambiguously show that the spatial self-alignment of the flake to the nanocavity facilitates light-matter coupling.

\section{Discussion}\label{DISCUSSION}

We have successfully fabricated hybrid nanocavities using hBN, WSe$_2$ and MoTe$_2$ which exhibit a range of $Q$ factor values. It is worth reiterating that the extracted $Q$ factors are $Q_\text{l}$ which is limited by either $Q_\text{c}$ and $Q_\text{i}$. The $Q_\text{i}$ is in turned determined only by the coupling losses to free space and is comprised of the theoretical ($Q_\text{th}$) and experimental ($Q_\text{ex}$) losses: $\frac{1}{Q_{\text{i}}} = \frac{1}{Q_{\text{th}}} + \frac{1}{Q_{\text{ex}}}$. The experimental $Q$ factors of the devices are lower than the theoretical values due to fluctuations in the fabricated structure parameters, for example, non-uniformity of air hole radii and lattice period, leading to scattering losses. The broken vertical symmetry of the device could also contribute to loss via the coupling of transverse electric and transverse magnetic modes~\cite{kuramochi2005,tanaka2006}. The irregular shape of the flake could also lead to losses as more of the flake edges overlap with the air holes~\cite{tomljenovic-hanic2007a}. Wrinkles, bubbles, and contaminants could also lead to air gaps between the 2D material flake and the PhC waveguide, affecting the cavity $Q$. Nonetheless, the surface roughness of the flakes are mostly smooth and the interface between the flake and the PhC substrate clean, especially for the WSe$_2$ devices which are fabricated using the anthracene-assisted transfer method~\cite{otsuka2021}. As such, the surface condition of the flake is likely not the main factor that affects the $Q$ factor. From the scanning electron micrograph of the fabricated PhC waveguide (Supplementary Fig. S7), the PhC air holes show imperfect roundness with slight ellipticity, which we believe to the main cause that limits the experimental $Q$ factors. 

Despite these issues, high $Q$ factors can be achieved in such 2D material-induced hybrid PhC nanocavities as demonstrated by our devices. The high $Q/V_\text{mode}$ of our hybrid nanocavity devices also enabled the observation of nonlinear effects such as the optical bistability at low excitation powers (Supplementary note 2 and Fig. S8). By comparing the experimentally observed guided mode redshift caused by the 2D materials and the required change in the slab refractive index of bare PhC waveguide to produce the same redshift in FDTD simulations, it is estimated that a monolayer flake results in ~0.1--0.2\% change in the local refractive index. This is consistent with previous reports which suggested that ultrahigh $Q$ PhC mode gap cavities could be formed in PhC waveguides with a local refractive index modulation as small as 0.1\%~\cite{notomi2008a, chiba2019}. Furthermore, the devices are stable, with the cavities showing no significant degradation even after being stored in ambient conditions for a time period ranging from several months to more than a year (see Supplementary Fig.~S9).

It is interesting to note that despite the 20 nm thickness of the hBN flake, the obtained experimental $Q$ is an order of magnitude higher than the theoretical $Q$ of a device with equally thick WSe$_2$ flake (Fig.~\ref{Fig1}d). The refractive index of  hBN is taken to be 2.2, smaller than that of the silicon PhC slab. In addition to modulating the local refractive index, the interface between the hBN flake and the PhC waveguide facilitates total internal reflection, resulting in good optical confinement and thus a high $Q$ factor. In contrast to hBN, TMDCs have refractive indices comparable or larger than that of silicon and thus do no facilitate total internal reflection at the flake/waveguide interface. Nonetheless, obtaining mono- and few-layer flakes with TMDCs is relatively easier.  This advantage helps overcome the larger refractive indices of TMDCs and enables the creation of high $Q$ hybrid nanocavities.

In conclusion, we have demonstrated experimentally that a 2D material flake can induce high $Q$ factor hybrid nanocavity in a PhC waveguide post-fabrication. The cavity can be formed at arbitrary location along the waveguide and could readily couple to the 2D material flake, forming a self-aligned device. It is worth emphasizing that even a monolayer flake is capable of forming a hybrid nanocavity, enabling the combination of the exceptional optical properties of monolayers with the flexibility of optical engineering offered by the PhC substrate. Our versatile approach to form hybrid PhC nanocavities and self-aligned devices can be extended to encompass a wide variety of 2D materials, allowing the creation of devices with diverse functionalities. By transferring flakes of suitable 2D materials, passive PhC waveguides can be transformed post-fabrication into active devices such as laser, modulator, and detector, facilitating the development of hybrid integrated photonic circuits.

\section{Methods}\label{METHODS}

\textit{Numerical simulations.} The photonic band structure is calculated using MIT Photonic Bands (MPB)~\cite{johnson2001}, considering a unit cell of the PhC waveguide with periodic boundary conditions in the $x$- and $y$-directions. The finite-difference time-domain (FDTD) simulations are performed using the open-source package MEEP~\cite{oskooi2010} on a computing cluster. The grid resolution is usually set to at least $a$/24 or higher, depending on the thickness of the simulated 2D materials. With subpixel averaging, we are able to simulate 2D material of thicknesses down to 5 nm and obtain reliable results. Numerical finite element method (FEM) simulations are carried out with COMSOL to simulate PhC structures with mono- to few-layer thick 2D materials. The simulated PhC waveguide consisting of 48 (14) air holes along the $\Gamma$ -- $K$ ($\Gamma$ -- $M$) direction is sufficiently large such that any further increase in size does not change the $Q$ factor, i.e., the $Q$ factor is limited only by out-of-plane losses. The type of simulated 2D materials is controlled by setting the value of the dielectric constant, obtained from reference~\citealt{laturia2018}. We assume a monolayer effective thickness of 0.7 nm in our simulations. Silicon has minimal absorption at the NIR regime where that cavity mode is. Aside from MoTe$_2$, hBN and WSe$_2$ are not expected to have strong absorption in the NIR and thus we do not include absorption losses in the simulations. We also performed simulations using the isotropic or anisotropic dielectric constants and found no significant difference in the results.

\textit{Silicon PhC waveguide fabrication.} The PhC waveguides are fabricated on a silicon-on-insulator substrate with a 200-nm-thick top silicon layer and a 1-$\mu$m-thick buried oxide layer. The PhC pattern is first defined on a resist mask by electron beam lithography, then the pattern is transferred onto the substrate via inductively coupled plasma using C$_{4}$F$_{8}$ and SF$_6$ gases. Following resist removal, the buried oxide layer is etched away using a solution of 20\% hydrofluoric acid to form air-suspended PhC waveguide structures. 

\textit{2D material dry transfer.} The hBN flakes (NIMS) are prepared on a polydimethylsiloxane (PDMS) sheet (Gelfilm by Gelpak) by mechanical exfoliation of bulk crystals. Suitable flakes are identified using an optical microscope and then transferred onto the target PhC waveguide using a homebuilt micromanipulator setup. MoTe$_2$ flakes (HQ Graphene) are prepared and transferred using the same method. 

WSe$_2$ flakes (HQ Graphene) are prepared on a commercially available 90-nm-thick SiO$_2$/Si substrates via mechanical exfoliation to enable the identification of the layer number via optical contrast. The WSe$_2$ flakes are then placed on the PhC waveguides using the anthracene-assisted transfer process~\cite{otsuka2021}. To grow the anthracene crystals, anthracene powder is heated to about 80$^\circ$C. The sublimated anthracene vapor will then recrystallize on the bottom surface of a glass slide placed at about 1 mm above the anthracene powder. The growth time is typically 10h. A small PDMS sheet is then placed on a glass slide, followed by an  anthracene crystal on the PDMS to form an anthracene/PDMS stamp. Next, this stamp is used to pick up the WSe$_2$ flake. The WSe$_2$ flake and the anthracene crystal were then transferred together onto the target PhC waveguide. Finally, the anthracene crystal is heated to about 80$^\circ$C or left in ambient conditions to sublime, leaving behind clean flakes. 

\textit{Optical spectroscopy.} PL measurements are performed with a homebuilt confocal microscopy system. A Ti:sapphire laser (Spectra Physics 3900S) is used for excitation, usually at a wavelength of 760$\--$780 nm. The excitation power is controlled using neutral density filters, while the polarization of the laser is adjusted using a half-wave plate to match the polarization of the cavity mode. The laser beam is focused on the samples using an objective lens (Olympus) of 50$\times$ magnification with a numerical aperture (NA) of 0.65. The emission from the sample is collected with the same objective lens, directed to a spectrometer (Princeton Instruments Acton SP2300), dispersed by a 150 lines/mm grating, and then detected with a liquid nitrogen-cooled InGaAs detector (Princeton Instruments  PyLoN IR). 

For transmission measurements, a wavelength tunable continuous-wave laser (Santec TSL-550) is used. A steering mirror and a 4$f$ system are used to displace the laser excitation spot while keeping the same detection spot. The light scattered from the sample is collected by the objective lens and coupled into an optical fiber to direct the signal to a photoreceiver (New Focus 2011).

For time-resolved measurements, a Ti:sapphire laser (Coherent MIRA) operating in pulsed mode is used for excitation. The excitation wavelength is set to 780 nm. The laser beam is focused on the sample with an objective lens (Olympus) of 100$\times$ magnification and NA of 0.85. The PL emission is collected by the same objective lens and then filtered spectrally with a long-pass filter to direct light of a specific wavelength range into an optical fiber connected to a superconducting nanowire single photon detector (Quantum Design Eos). All measurements are carried out at room temperature. The samples are kept in a nitrogen gas environment in bid to reduce the oxidation rate of the TMDC flakes.

\begin{acknowledgments}
This work is supported in part by JSPS KAKENHI (JP22K14623, JP20H02558, JP20J00817, JP22K14624, JP22K14625, JP22F22350 and JP23H00262) and ARIM of MEXT (JPMXP1222UT1138), as well as the RIKEN Incentive Research Projects. K.W. and T.T. acknowledge support from the JSPS KAKENHI (Grant Numbers JP21H05233 and JP23H02052) and World Premier International Research Center Initiative (WPI), MEXT, Japan. C.F.F. is supported by the RIKEN SPDR fellowship. Y.-R.C. is supported by the JSPS Postdoctoral Fellowship. We acknowledge support by the RIKEN Information Systems Division for the use of the Supercomputer HOKUSAI BigWaterfall and SailingShip, as well as the use of the COMSOL license.\\
\end{acknowledgments}

\section*{Author contributions}
C.F.F conceived the idea and performed the numerical simulations. C.F.F fabricated the PhC waveguide under the guidance of D.Y. C.F.F carried out the 2D materials transfer with assistance from N.F and Y.-R.C. C.F.F performed the optical measurements with some assistance from D.Y and S.F. K.W and T.T provided the bulk hBN crystals. C.F.F analyzed the data and wrote the manuscript. All authors contributed to the discussion of the results and the manuscript. Y.K.K supervised the project.

\end{document}